\documentstyle[11pt]{amsart}

\begin{document}

\title[Combinatorial Identities and Quantum State Densities]
{Combinatorial Identities and Quantum State Densities of
Supersymmetric Sigma Models on N-Folds}

\author{M. C. B. Abdalla}
\address{Instituto de F\'{\i}sica Te\'orica, Universidade Estadual Paulista,
Rua Pamplona 145, 01405-900 - S\~ao Paulo, SP, Brazil\,\, {\em
E-mail address:} {\rm mabdalla@@ift.unesp.br}}
\author{A. A. Bytsenko}
\address{Departamento de F\'{\i}sica, Universidade Estadual de Londrina,
Caixa Postal 6001, Londrina, PR, Brazil \,\, {\em E-mail address:}
{\rm abyts@@uel.br}}
\author{M. E. X. Guimar\~aes}
\address{Departamento de Matem\'atica, Universidade de
Bras\'{\i}lia, Campus Universit\'ario, Bras\'{\i}lia, DF,
Brazil\,\, {\em E-mail address:} {\rm marg@@unb.br}}

\maketitle

\begin{abstract}

There is a remarkable connection between the number of quantum
states of conformal theories and the sequence of dimensions of Lie
algebras. In this paper, we explore this connection by computing
the asymptotic expansion of the elliptic genus and the microscopic
entropy of black holes associated with (supersymmetric) sigma
models. The new features of these results are the appearance of
correct prefactors in the state density expansion and in the
coefficient of the logarithmic correction to the entropy.

\end{abstract}

\section{Introduction}

The combinatoric identities with which we shall be concerned play
an important role in a number of physical models. In particular, such
identities can be associated with the elliptic genus partition
function of supersymmetric sigma models on the $N-$folds and play
a special role in string and black hole dynamics. Before entering into
the specific problem, we would like to introduce the reader to some
formal aspects found in the  mathematical literature
\cite{Fuks,Kac}, giving two examples in string theory which
we shall use later on.
\newline
\newline
{\bf $CY_3-$folds}. First we consider type IIA string theory
compactified on Calabi-Yau three folds $CY_3$. Recently, black holes
have been studied in the $N=2$ supergravity corresponding to type
IIA strings. This theory can also be viewed as M-theory on
$CY_3\times S^1$ and extremal black holes are microscopically
represented by fivebranes wrapping on $P\times S^1$, where $P$ is a
four-cycle, $P\subset CY_3$. The microscopic entropy of the
fivebrane has been determined from a two-dimensional (0,4) sigma
model, whose target space includes the fivebrane moduli space
\cite{Maldacena1}. We are interested in the $CY_3$ geometry. The
charge forms are in one to one correspondence with the elements of
$H^q(CY_3, {\mathcal L}\otimes \Omega^p)$, where ${\mathcal L}$ is
the line bundle for which $c_1({\mathcal L})=[F_{CY}]$;
\,$c_1({\mathcal L})$ is the first Chern class of ${\mathcal L}$.
$H^q(CY_3, {\mathcal L}\otimes \Omega^p)$ vanishes for $q>0$ and
large $c_1({\mathcal L})$. One can compute $\dim H^0(CY_3, {\mathcal
L}\otimes \Omega^p)$ using the Riemann-Roch formula, the result is
(see also \cite{Gaiotto}):
\begin{eqnarray}
h_0 & = & {\rm dim}\, H^0(CY_3, {\mathcal L}) =
\int \left[ \frac{F_{CY}^3}{6}+\frac{c_2\wedge F_{CY}}{12} \right]
= {\mathcal D} + \frac{1}{12}c_2\cdot P,
\nonumber \\
h_1 & = & {\rm dim}\, H^0(CY_3, {\mathcal L}\otimes \Omega^1) =
\int \left[
\frac{F_{CY}^3}{2}-\frac{3c_2\wedge F_{CY}}{4}+\frac{c_3}{2}\right]
= 3{\mathcal D} -\frac{3}{4}c_2\cdot P-\frac{\chi}{2},
\nonumber \\
h_2 & = & {\rm dim}\, H^0(CY_3, {\mathcal L}\otimes \Omega^2) =
\int \left[
\frac{F_{CY}^3}{2}-\frac{3c_2\wedge F_{CY}}{4}-\frac{c_3}{2}
\right] = 3{\mathcal D} -\frac{3}{4}c_2\cdot P+\frac{\chi}{2},
\nonumber \\
h_3 & = & {\rm dim}\, H^0(CY_3, {\mathcal L}\otimes \Omega^3) = h_0
\mbox{.}
\end{eqnarray}
Here $c_k$ are the k-th Chern classes, $\chi$ is the Euler
characteristic and $P$ is a four-cycle of a manifold (for more
notations and details see for example \cite{Maldacena1,Gaiotto}).

Our goal is to compute a partition function $ Z(q)\simeq {\rm
Tr}\, [{\mathcal O}q^N] $ in a black hole geometry, where the
trace is calculated over the multi-brane Hilbert space and
$\mathcal O$ is an appropriate operator insertion. We have to
count the multiparticle primaries choosing a basis of states.
Actually the counting of configurations is in one to one
correspondence to the counting of states for conformal field theory
with $\sum_{({\rm even}\,\,j)}h_j$ bosons and $\sum_{({\rm
odd}\,\,j)}h_j$ fermions and total momentum $N$. The number of
states would correspond to the coefficient of $D(n)$ in the
expansion of the generating function
\begin{equation}
Z={\rm Tr}q^N =
\prod_n \frac{(1+q^n)^{\sum_{({\rm odd}\,\,j)}h_j}}
{(1-q^n)^{\sum_{({\rm even}\,\,j)}h_j}}
= \sum_n{D}(n)q^n
\mbox{.}
\label{generating}
\end{equation}
{\bf Supersymmetric sigma model on $N-$folds}. Next example is a
sigma model on the $N-$fold symmetric product $S^NX$ of a K\"{a}hler
manifold $X$, which is the $S^N X = X^N/S_N$ orbifold space. $S_N$
is the symmetric group of $N$ elements. The Hilbert space of an
orbifold field theory can be decomposed into twisted sectors
${\mathcal H}_{\gamma}$, that are labelled by the conjugacy classes
$\{\gamma\} $ of the orbifold group $S_N$ \cite{Dixon2,Dijkgraaf1}.
For a given twisted sector one can keep the states invariant under
the centralizer subgroup $\Gamma_{\gamma}$ related to the element
$\gamma$. Let ${\mathcal H}_{\gamma}^{\Gamma_{\gamma}}$ be an
invariant subspace associated with $\Gamma_{\gamma}$; the total
orbifold Hilbert space takes the form ${\mathcal H}(S^N X)=
\oplus_{\{\gamma\}} {\mathcal H}_{\gamma}^{\Gamma_{\gamma}}$. One
can compute the conjugacy classes $\{\gamma\}$ by using a set of
partitions $\{N_n\}$ of $N$, namely $\sum_nnN_n=N$, where $N_n$ is
the multiplicity of the cyclic permutation $(n)$ of $n$ elements in
the decomposition of $\gamma$: $\{\gamma\}=\sum_{j=1}(j)^{N_j}$. For
this conjugacy class the centralizer subgroup of a permutation
$\gamma$ is $\Gamma_{\gamma}=S_{N_1}\otimes_{j=2}(S_{N_j}>\!\!\!\lhd
{Z} _j^{N_j})$ \cite{Dijkgraaf1}, where each subfactor $S_{N_n}$ and
${Z}_n$ permutes the $N_n$ cycles $(n)$ and acts within one cycle
$(n)$ correspondingly. Following the lines of \cite{Dijkgraaf1} we
may decompose each twisted sector ${\mathcal
H}_{\gamma}^{\Gamma_{\gamma}}$ into a product over the subfactors
$(n)$ of $N_n-$fold symmetric tensor products, ${\mathcal
H}_{\gamma}^{\Gamma_{\gamma}}= \otimes_{n>0}S^{N_n}{\mathcal
H}_{(n)}^{Z_{n}}$, where $S^N{\mathcal H}\equiv (\otimes^N{\mathcal
H})^{S_N}$.

It has been shown that the partition function for (sub)Hilbert
space of a supersymmetric sigma model coincides with the elliptic
genus \cite{Landweber}. If $\chi({\mathcal H}_{(n)}^{{Z}_n};q,y)$
admits the extension $\chi({\mathcal H};q,y)=\sum_{m\geq
0,\ell}G(nm,\ell)q^my^{\ell}$, the following result holds (see
\cite{Kawai,Dijkgraaf1,Dijkgraaf2}):
\begin{eqnarray}
\prod_{m\geq 0,\ell}
\left(1-pq^my^{\ell}\right)^{-G(nm,\ell)}
& = & \sum_{N\geq 0}p^N\chi(S^N{\mathcal H}_{(n)}^{{\mathbb Z}_n};q,y),
\nonumber \\
W(p; q, y) = \prod_{n>0,m\geq 0,\ell}
\left(1-p^nq^my^{\ell}\right)^{-G(nm,\ell)}
& = & \sum_{N\geq 0}p^N\chi(S^NX;q,y)
\mbox{,}
\label{Id2}
\end{eqnarray}
$p={\bf e}[\rho]$, $q ={\bf e}[\tau]$, $y={\bf e}[z]$, and ${\bf e}[x]
\equiv \exp [2\pi i x]$.
Here $\rho$ and $\tau$ determine the complexified
K\"{a}hler form and complex structure modulos of ${\mathbb T}^2$
respectively,
and $z$ parametrizes the $U(1)$ bundle on ${\mathbb T}^2$.
The Narain duality
group $SO(3,2,{\mathbb Z})$ is isomorphic to the Siegel modular
group $Sp(4, {\mathbb Z})$ and it is convenient to combine
the parameters $\rho, \tau$ and a
Wilson line modules $z$ into a $2\times 2$ matrix belonging to the
Siegel upper half-plane of genus two,
$\Xi = \left(
\begin{array}{ll}
\rho\,\,\,z &  \\
z\,\,\, \tau &
\end{array}
\right)$, with $\Im \rho >0,\, \Im \tau >0$, ${\rm det}\,\Im \Xi
>0$. The group $Sp(4, {\mathbb Z}) \cong SO(3,2,{\mathbb Z})$ acts
on the $\Xi$ matrix by fractional linear transformations $\Xi
\rightarrow (A\Xi +B)(C\Xi + D)^{-1}$. Note that for a Calabi-Yau
space the $\chi-$genus is a weak Jacobi form of zero weight and
index $d/2$ \cite{Hirzebruch1}. For $q=0$ the elliptic genus reduces
to a weighted sum over the Hodge numbers, namely $\chi(X;0,y)=
\sum_{j,k}(-1)^{j+k}y^{j-\frac{d}{2}}h^{j,k}(X)$. For the trivial
line bundle the symmetric product $\prod_{n>0}
\left(1-p^n\right)^{-\chi(X)}$ (see Section 3 for details) can be
associated with the simple partition function of a second quantized
string theory.

The paper is organized as follows. In Section 2 we discuss the
homological method of the relationship between Lie algebras and
combinatorial identities following the lines of
\cite{Garland,Fuks}. The asymptotic expansion of the elliptic
genus and the microscopic entropy of a black hole associated with
a supersymmetric sigma model are given in Section 3. The problem of
microscopically computing the entropy of black holes has
already been solved. In the present paper the new features are the
appearance of correct prefactors in the state density expansion
and in the coefficient of the logarithmic correction to the
entropy. The fact that the computation of the number of states for
conformal theories, and therefore computing the entropy of black
holes, would connected to the sequence of dimensions of Lie
algebras is quite intriguing and we finish this paper with a
discussion on this point.

\section{Combinatorial identities and graded Lie algebras}

One of the universal methods of obtaining combinatorial identities
of (\ref{generating}), (\ref{Id2}) types is the
Euler-Poincar{\'e} formula associated with a complex consisting of
finite-dimensional linear spaces. In this section we briefly
discuss the homological aspects of identities following the lines of
\cite{Garland,Fuks}. Our remarks here are designed to provide the
readers with a brief introduction to these identities and to
indicate how it could be derived from results of (graded) Lie
algebras. The relationship between combinatorial identities and
Lie algebras was discovered by Macdonald \cite{Macdonald} (on the
whole all combinatorial identities are related to Lie algebras).
In this Section we shall apply Euler-Poincar{\'e} formula to chain
complexes of Lie algebras.

Let ${\frak g}$ be a finite-dimensional Lie algebra and let
$C_q({\frak g})$ be the space of $q-$dimensional chain of ${\frak
g}$. The Euler-Poincar{\'e} formula gives
\begin{eqnarray}
&& \sum_q(-1)^qc_q^{(m)} = \sum_q(-1)^qh_q^{(m)},
\,\,\,\,\,\,\,\, m\in {\mathbb N},
\nonumber \\
&& c_q^{(m)} = {\rm dim}\,C_{q}^{(m)}({\frak
g}),\,\,\,\,\,\,\,\,\,\, h_q^{(m)} = {\rm dim}\,H_{q}^{(m)}({\frak
g}) \mbox{,}
\end{eqnarray}
where $H_{q}({\frak g})$ is the homology of the complex
$\{C_q({\frak g})\}$.

Introducing the $x$ variable we can rewrite this sequence of
identities as a formal power series:
\begin{equation}
\sum_{q,m}(-1)^qc_q^{(m)}x^m = \sum_{q,m}(-1)^qh_q^{(m)}x^m
=\prod_j(1-x^j)^{{\rm dim}\,{\frak g}_{(j)}}
\mbox{.}
\label{EP}
\end{equation}
Therefore, in order to get the identity in its final form the
homology $H_q^{(m)}({\frak g})$ has to be computed.

Let us suppose that the Lie algebra ${\frak g}$ possesses a
polygrading ${\frak g} = \oplus {\frak g}_{(m_1, ..., m_k)}$. The
following result holds:

{\bf Theorem 1} (D. B. Fuks \cite{Fuks}, Theorem 3.2.3) \,\,\, Let
$ {\frak g} = \oplus_{\scriptstyle m_1\geq 0, ..., m_k\geq 0
\atop\scriptstyle m_1+...+m_k>0} {\frak
g}_{(m_1, ..., m_k)} 
$ be the\,\,  (poly)graded \,\, Lie algebra satisfying \,\,\, $
{\rm dim}\, {\frak g}_{(m_1, ..., m_k)} < \infty . $ \,\, If $
{\rm dim}\,H_q^{(m_1,...,m_k)} = h_q^{(m_1,...,m_k)}, $ the formal
power series in $x_1,...,x_k$ satisfy the following identity
\begin{equation}
\label{fuks}
\prod_{j_1,...,j_k}\left(1-x_1^{j_1}...x_k^{j_k}\right)^ {{\rm
dim}\, {\frak g}_{(m_1, ..., m_k)}}
=\sum_{q,m_1,...,m_k}(-1)^qh_q^{m_1,...,m_k}x_1^{m_1}...x_k^{m_k}
\mbox{.}
\end{equation}

We can apply Theorem 1 to compute the Lie algebras homology which
has been carried out in the papers \cite{Feigin,Lepowsky}. Assume
that a Hermitian or Euclidean metric can be choosen in every space
${\frak g}_{(\lambda)}$. As a consequence, all the space
$C^q_{(\lambda)}({\frak g})$ acquire a metric and one can identify
$C^q_{(\lambda)}({\frak g})$ with $[C^q_{(\lambda)}({\frak g})]'$,
i.e. with $C_q^{(\lambda)}({\frak g})$. The homology of  the Lie
algebras can be entirely computed by using the Laplace operator
${\frak L}^q_{(\lambda)}$. Every element of the space
$H^q_{(\lambda)}({\frak g})$ can be represented by an unique
harmonic cocycle from $C^q_{(\lambda)}({\frak g})$ (see, for
example, \cite{Fuks}). It means that there is a natural
isomorphism:
\begin{equation}
{\rm Ker}\, {\frak L}^q_{(\lambda)} =
H^q_{(\lambda)}({\frak g})
\mbox{.}
\end{equation}

As an example, let us briefly consider an application of Theorem 1
to graded Lie subalgebras $N^{+}({\frak g}^A)$ of Kac-Moody
algebras. The algebra $N^{+}({\frak g}^A)$ is constructed from the
Cartan matrix $A$. The matrix $A = ||a_{ij}||_{i,j=1}^n$ is square
integer, $a_{ij}=2,\, i,j=1,2,...,n,\, a_{ij}\leq 0$ for $i\neq
j$, and for which there exist positive numbers $b_1, ..., b_n$,
such that the matrix $bA = ||b_ia_{ij}||$ is symmetric. The
Kac-Moody algebra ${\frak g}^A$ with Cartan matrix $A$ is the
complex Lie algebra with generators $e_1,...,e_n$. We can
construct ${\mathbb Z}^n-$grading in the Kac-Moody algebra ${\frak
g}^A$, choosing the appropriate form of $\{{\rm
deg}\,e_j\}_{j=1}^n$. The following theorem occurs:

{\bf Theorem 2} (D. B. Fuks \cite{Fuks}) \,\,\, Formula
(\ref{fuks}) applied to graded Kac-Moody algebras gives:
\begin{equation}
\prod_{\scriptstyle k_1\geq 0, ..., k_n\geq 0
\atop\scriptstyle k_1+...+k_n>0}
(1-x_1^{k_1}...x_n^{k_n})^{{\rm dim}\,{\frak g}_{(k_1,...,k_n)}^A}
=\sum_{Q(j_1,...,j_n)=0}L(j_1,...,j_n)x_1^{j_1}...x_n^{j_n}
\mbox{,}
\end{equation}
where $L(j_1,...,j_n)$ are certain coefficients.

The identities which correspond to the $N^{+}({\frak g}^A)$
algebra with Cartan matrices of rank $n-1$ with negative
eigenvalues are usually called Macdonald identities. Combinatorial
identities related to Kac-Moody algebras with other Cartan
matrices are also of interest (see next section).

\section{Asymptotics of generating functions}

One can apply the Theorem 1 to the ${\mathbb Z}^n-$grading of the
${\frak s}{\frak l}(n,{\mathbb C})$ subalgebras of Lie algebras
together with the Theorem 2. It gives the Macdonald identities
series containing the Gauss-Jacobi identity (see for details
\cite{Fuks}):
\[
\prod_{m=1}^{\infty}(1-x_1^mx_2^m)(1-x_1^mx_2^{m-1})(1-x_1^{m-1}x_2^m)
= 1 +\sum_{k=1}^{\infty}(-1)^k ( x_1^{\frac{k(k+1)}{2}}
x_2^{\frac{k(k-1)}{2}} +
x_1^{\frac{k(k-1)}{2}}x_2^{\frac{k(k+1)}{2}} ). \] We can make use
of trivial transformations of the Gauss-Jacobi identity. It
reduces to the series
\begin{equation}
\prod_{n=1}^{\infty}(1-x^n)^3 =\sum_{k=1}^{\infty}(-1)^{k-1}
(2k-1)x^{\frac{k(k-1)}{2}}
\mbox{,}
\end{equation}
which is the cube of the ``Euler function''
$\prod_{n=1}^{\infty}(1-x^n)$. Interesting combinatorial
identities may be obtained by applying the Theorem 1 to graded Lie
algebras, but only for some chosen values of the power $k$ can the
function $\prod_{n=1}^{\infty}(1-x^n)^k$ be presented by a power
of Euler function.  Some arguments on ``distinguished powers''
(which is in correspondence to the sequence of dimensions of Lie
algebras ) the reader can find in \cite{Fuks1} (subsections 3 and
4).

Since the coefficient $D(n)$ in the expansion of the generating
function in its final form is not always known, we shall simplify
the calculations and apply its asymptotic limit. The
multi-component version of the classical generating functions has
the form
\begin{equation}
{\mathfrak G}_{\pm}(z)=\prod_{{\bf n}\in {\mathbb Z}^p/\{{\bf 0}\}}
\left[1\pm
\exp\left(-z\omega_{{\bf n}}({\bf a},
{\bf g})\right)\right]^{\pm \sigma}
\mbox{,}
\end{equation}
where $\Re z>0$,\, $\sigma>0$, $\omega_{{\bf n}}({\bf a},
{\bf g})$ is given by
$
\omega_{\bf n}({\bf a}, {\bf g})=
\left(\sum_{\ell}a_\ell(n_\ell+{\rm g}_\ell)^2\right)^{1/2}
$,
${\rm g}_\ell$, and $a_\ell$ are some real numbers
(for arbitrary spectral forms $\omega_{\bf n}^2$ see, for example,
\cite{Bytsenko030}).
The total number of quantum states can be described by the
functions $D_{\pm}(n)$ defined by
$
{\mathcal K}_{\pm}(t)=\sum_{n=0}^{\infty}D_{\pm}(n)t^n
\equiv {\mathfrak G}_{\pm}(-\log t),
$
where $t<1$, and $n$ is a total quantum number. The $p-$dimensional
Epstein zeta function
$Z_p\left|_{\bf f}^{\bf g}\right|(z,\varphi)$
associated with the quadratic
form $\varphi [{\bf a}({\bf n}+{\bf g})]=
(\omega_{\bf n}({\bf a}, {\bf g}))^2$
for $\Re\,z>p$ is given by the formula
\begin{equation}
Z_p\left| \begin{array}{ll}
{\rm g}_1\,...\,{\rm g}_p \\
f_1\,...\,f_p\\
\end{array} \right|(z,\varphi)=\sum_{{\bf n}\in {\Bbb Z}^p}{}'
\left(\varphi[{\bf a}({\bf n}+{\bf g})]\right)^{-\frac{z}{2}}
e^{2\pi i({\bf n},{\bf f})}
\mbox{,}
\label{Epstein}
\end{equation}
where $({\bf n},{\bf f})=\sum_{i=1}^p n_if_i$,\, $f_i$ are real
numbers and the prime on $\sum {'}$ means to omit the term ${\bf
n} =-{\rm {\bf g}}$ if all the ${\rm g}_i$ are integers. For $\Re
z<p$,\, $Z_p\left|_{\bf f}^{\bf g} \right|(z,\varphi)$ is
understood to be the analytic continuation of the right hand side
of  Eq. (\ref{Epstein}). Note that $Z_p\left|_{\bf f}^{\bf
g}\right|(z,\varphi)$ is an entire function in the complex
$z-$plane except for the case when all the $f_i$ are integers. In
this case $Z_p\left|_{\bf f}^{\bf g}\right|(z,\varphi)$ has a
simple pole at $z=p$ with residue $ A(p)= 2\pi^{p/2}[({\rm
det}\,{\bf a})^{1/2}\Gamma(p/2)]^{-1} $, which does not depend on
the winding numbers ${\rm g}_\ell$.

By means of the asymptotic expansion of ${\mathfrak G}_{\pm}(z)$ for
small $z$, one arrives at a complete asymptotic limit of
$D_{\pm}(n)$ \cite{Bytsenko93,Elizalde94,Bytsenko96}:
\begin{eqnarray}
D_{\pm}(n)_{n\rightarrow \infty} & = &
{\mathcal C}_{\pm}(p)n^{\frac{2\sigma
Z_p\left|_{{\bf 0}}^{{\bf g}}\right|
(0,\varphi)-p-2}{2(1+p)}}
\exp\left\{\frac{1+p}{p}[\sigma A(p)
\Gamma(1+p)\zeta_{\pm}(1+p)]^
{\frac{1}{1+p}}n^{\frac{p}{1+p}}\right\} \nonumber \\
& \times & [1+{\mathcal O}(n^{-k_{\pm}})]
\mbox{,}
\label{asym1}
\end{eqnarray}
\begin{equation}
{\mathcal C}_{\pm}(p)=[\sigma A(p)\Gamma(1+p)
\zeta_{\pm}(1+p)]^{\frac{1-2\sigma
Z_p\left|_{{\bf 0}}^{{\bf g}}\right|(0,\varphi)}{2p+2}}
\frac{\exp\left[\sigma (d/dz)
Z_p\left|_{{\bf 0}}^{{\bf g}}\right|(z,\varphi)|_{(z=0)}\right]}
{[2\pi(1+p)]^{1/2}}
\mbox{,}
\label{asym2}
\end{equation}
\newline
where $\zeta_{-}(z)\equiv \zeta_R(z)$ is the Riemann zeta function,
$\zeta_{+}(z) = (1-2^{1-z})\zeta_R(z)$,
$
k_{\pm}= p/(1+p)\min \left({\mathcal C}_{\pm}(p)/p-\delta/4,
1/2-\delta\right)
$, and $0<\delta<2/3$.

\subsection{Asymptotics of the elliptic genus}

If $y={\bf e}[z] =1 $ then the elliptic genus degenerates to the
Euler number or Witten index \cite{Hirzebruch2}. For the symmetric
product this gives the following identity
\begin{equation}
W(p) = \sum_{N\geq 0}p^N\chi(S^NX)=
\prod_{n>0} \left(1-p^n\right)^{-\chi(X)}
\mbox{.}
\label{character}
\end{equation}
Thus this character is almost a modular form of weight $-\chi(X)/2$.
Eq. (\ref{character}) is similar to the denominator formula of a
(generalized) Kac-Moody algebra \cite{Borcherds}. A denominator
formula can be written as follows:
\begin{equation}
\sum_{\sigma\in {\mathcal W}}\left({\rm sgn}(\sigma)\right)e^{\sigma(v)}
=e^{v}\prod_{r>0}\left(1-e^{r}\right)^{{\rm mult}(r)}
\mbox{,}
\label{denom}
\end{equation}
where $v$ is the Weyl vector, the sum on the left hand side is over all
elements of the Weyl group ${\mathcal W}$, the product on the right
hand side runs
over all positive roots (one has the usual notation of root spaces,
positive
roots, simple roots and Weyl group, associated with Kac-Moody algebra)
and
each term is weighted by the root multiplicity
${\rm mult}(r)$. For the
$su(2)$ level, for example, an affine Lie algebra (\ref{denom})
is just the Jacobi triple product identity. For generalized Kac-Moody
algebras there is the following denominator formula:
\begin{equation}
\sum_{\sigma\in {\mathcal W}}\left({\rm sgn}(\sigma)\right)\sigma
(e^{v}\sum_{r} \varepsilon(r)e^{r})=e^{v}\prod_{r>0}
\left(1-e^{r} \right)^{{\rm mult}(r)}
\mbox{,}
\label{denom}
\end{equation}
where the correction factor on the left hand side involves
$\varepsilon(r)$ which is $(-1)^n$ if $r$ is the sum of $n$
distinct pairwise orthogonal imaginary roots and zero otherwise.

The logarithm of the partition function $W(p;q,y)$ is the one-loop free
energy $F(p;q,y)$ for a string on ${\mathbb T}^2\times X$:
\begin{eqnarray}
\!\!\!\!\!\!\!\!\!\!\!\!\!\!\!
F(p;q,y) & = & \mbox{log}W(p;q,y)=-\sum_{n>0,m,\ell}G(nm,\ell)\mbox{log}
\left(1-p^nq^my^{\ell}\right)
\nonumber \\
& = & \!\!\!\! \sum_{n>0,m,\ell,k>0}
\frac{1}{k}G(nm,\ell)p^{kn}q^{km}y^{k\ell}
=\sum_{N>0}p^N\!\!\sum_{kn=N}\frac{1}{k}
\sum_{m,\ell}G(nm,\ell)q^{km}y^{k\ell}
\mbox{.}
\end{eqnarray}
The free energy can be written as a sum of Hecke operators $T_N$
acting on the elliptic genus of $X$
\cite{Borcherds,Gritsenko,Dijkgraaf1}: $
F(p;q,y)=\sum_{N>0}p^NT_N\chi(X;q,y) $.

The goal now is to calculate an asymptotic expansion
of the elliptic genus
$\chi(S^NX;q,y)$. The degeneracies for the sigma model are given by the
Laurent inversion formula:
$
\chi(S^NX;q,y)= (2\pi i)^{-1}\oint W(p,q,y) p^{-N-1}dp,
$
where the contour integral is taken on a small circle around the origin.
Let the Dirichlet series
\begin{equation}
{D}(s;\tau,z)=
\sum_{(n,m,\ell)>0}\sum_{k=1}^{\infty}\frac{{\bf e} [\tau
mk+z\ell k]G(nm,\ell)}{n^sk^{s+1}}
\mbox{}
\label{Dir}
\end{equation}
converge for $0<\Re\,s<\alpha$. We assume that series (\ref{Dir}) can be
analytically continued in the region $\Re\,s\geq-C_0\,\,(0<C_0<1)$
where it
is analytic excepting a pole of order one at $s=0$ and $s=\alpha$, with
residue $\mbox{Res}[{D}(0;\tau,z)]$ and
$\mbox{Res}[{D}(\alpha;\tau,z)]$ respectively. Besides, let
${D}(s;\tau,z)={\mathcal O}(|\Im\,s|^{C_1})$ uniformly in
$\Re\,s\geq-C_0$ as $|\Im\,s|\rightarrow\infty $,
where $C_1$ is a fixed positive real number. The Mellin-Barnes
representation of the function $F(t;\tau,z)$ has the form
\begin{equation}
{\hat M}[F](t;\tau,z) =
\frac{1}{2\pi i}\int_{\Re\,s=1+\alpha}t^{-s}\Gamma(s)
{D}(s;\tau,z)ds \mbox{.}  \label{Mellin}
\end{equation}
The integrand in Eq. (\ref{Mellin}) has a first order pole at
$s=\alpha$ and
a second order pole at $s=0$. Shifting the vertical contour from
$\Re\,s=1+\alpha$ to $\Re\,s=-C_0$ (this procedure is permissible)
and making use of the residues theorem one obtains
\begin{eqnarray}
F(t;\tau,z) & = & t^{-\alpha}\Gamma(\alpha){\rm Res}
[D (\alpha;\tau,z)]
+\lim_{s\rightarrow 0}\frac{d}{ds}[s{D}(s;\tau,z)]
\nonumber \\
& - & (\gamma+{\rm log}\,t){\rm Res}[{D}(0;\tau,z)]
+\frac{1}{2\pi i}
\int_{\Re\,s=-C_0}\!\!t^{-s}\Gamma(s){D}(s;\tau,z)ds
\mbox{,}
\label{F}
\end{eqnarray}
where $t\equiv 2\pi(\Im\rho-i\Re\rho)$.
The absolute value of the integral
in (\ref{F}) can be estimated to behave as
${\mathcal O}\left((2\pi\Im\,\rho)^{C_0}\right)$.

In the half-plane $\Re t>0$ there exists an asymptotic expansion for
$W(t;\tau,z)$ uniformly in
$|\Re\rho|$ for $|\Im\rho|\rightarrow 0, \,\, |
\mbox{arg}(2\pi i\rho)|\leq \pi/4,\,\,|\Re\rho|\leq 1/2$
and given by
\begin{eqnarray}
{\rm log}\,W(t;\tau,z) & = & {\rm Res}[{D}
(\alpha;\tau,z)]\Gamma(\alpha)t^{-\alpha} -{\rm
Res}[{D}(0;\tau,z)]{\rm {log}t}
\nonumber \\
& - & \gamma {\rm Res}[{D}(0;\tau,z)]
+\lim_{s\rightarrow 0}
\frac{d}{ds}[s{D}(s;\tau,z)]+
{\mathcal O}\left( |2\pi\Im\tau|^{C_0}\right)
\mbox{.}
\end{eqnarray}
Repeating the above, we obtain the result (\ref{asym1}), (\ref{asym2})
with obvious modifications:
$\{n;p\} \Longrightarrow \{N; \alpha\},
$
$
D_{\pm}(n) \Longrightarrow \chi(S^NX;\tau,z),
$
and
\begin{eqnarray}
\sigma Z_p\left|_{{\bf 0}}^{{\bf g}}\right|(0,\varphi)
& \Longrightarrow & {\rm Res} [{D}(0;\tau,z)],
\nonumber \\
\sigma A(p)
\zeta_{\pm}(1+p)\Gamma(1+p) & \Longrightarrow &
{\rm Res}
[{D} (\alpha;\tau,z)]\Gamma(1+\alpha),
\nonumber \\
\sigma \lim_{z\rightarrow 0}
\frac{d}{dz}
Z_p\left|_{{\bf 0}}^{{\bf g}}\right|(z,\varphi)
& \Longrightarrow &
\lim_{s\rightarrow 0}
\frac{d}{ds} [s{D}(0;\tau,z)]-\gamma{\rm Res}
[{D}(0;\tau,z)]
\mbox{,}
\end{eqnarray}
where $\gamma$ is the Euler constant. These results have an universal
character for all elliptic genera associated to Calabi-Yau spaces.
\newline
\newline
{\bf Note}. We go into some facts related to orbifoldized
elliptic genus of $N=2$ superconformal field theory.
The contribution of the untwisted sector to the
orbifoldized elliptic genus is the function
$\chi(X;\tau,z)\equiv \Phi_{00}(\tau, z)$,
whereas
\begin{equation}
\phi\left(\frac{a\tau+b}{c\tau+d},\frac{z}{c\tau+d}\right)=
\Phi_{00}(\tau,z){\bf e}\left[\frac{rcz^2}{
c\tau+d}\right], \,\,\,\left(
\begin{array}{ll}
a\,\,\,b &  \\
c\,\,\,d &
\end{array}
\right)\in SL(2,{\mathbb Z})
\mbox{,}
\end{equation}
$r=d/2$. The contribution of the twisted $\mu-$sector projected
by $\nu$ is \cite{Kawai}:
$
\Phi_{\mu\nu}(\tau,z)=\Phi_{00}(\tau,z+\mu\tau+\nu){\bf e}
\left[d(\mu\nu+\mu^2\tau+2\mu z)/2\right],\mu,\nu\in {\mathbb Z}.
$
For some suitable integers $P$ and $\ell$ the orbifoldized elliptic
genus can be defined by
\begin{equation}
\phi(\tau,z)_{{\rm orb}}\stackrel{def}{=}\frac{1}{\ell}\sum_{\mu,
\nu=0}^{\ell -1}(-1)^ {P(\mu+\nu+\mu\nu)}
\Phi_{\mu\nu}(\tau,z)
\mbox{.}
\end{equation}
Using the transformation properties of the function
$\Phi_{\mu\nu}(\tau,z)$ one can obtain the asymptotic expansion
for the orbifoldized elliptic genus. In fact we can introduce a
procedure, starting with the expansion of the elliptic genus of
the untwisted sector, to compute the asymptotics of the elliptic
genus of the twisted sector.

\subsection{The microscopic entropy}

Results of the previous sections can be used to calculate the
ground state degeneracy of systems with quantum numbers of certain
states of extreme black holes. We can compute asymptotics of
functions ${\mathfrak G}_{\pm}(z), \, \chi(S^NX;\tau,z)$
associated with a gas of species of massless quanta. In the
context of superstring dynamics, for example, the asymptotic state
density gives a precise computation of the entropy of a black
hole. The black hole entropy ${S}(N)$ becomes
\begin{eqnarray}
{S}(N) & = & {\rm log}\, \chi(S^NX;\tau,z) \simeq
{S}_0 + {\mathcal A}(\alpha){\rm log} (S_0)
+ ({\rm Const.}),
\nonumber \\
{\mathcal A}(\alpha) & = &
(2\alpha)^{-1}\{2{\rm Res}[{D}(0;\tau,z)]-
2-\alpha\}
\mbox{.}
\label{entropy}
\end{eqnarray}
The leading term in Eq. (\ref{entropy}) has the form
\begin{equation}
S_0 = {B}(\alpha)N^{\alpha/(1+\alpha)},\,\,\,\,\,\,\,
{B}(\alpha) =
\frac{1+\alpha}{\alpha}\left\{{\rm Res}[{D}
(\alpha;\tau,z)]
\Gamma(1+\alpha)\right\}^{1/(1+\alpha)}
\mbox{.}
\end{equation}
${\mathcal A}(\alpha)$ is the coefficient of the logarithmic
correction to the entropy, it depends on the complex
dimension $d$ of a K\"{a}hler manifold $X$.

In conclusion, we note that the asymptotic state density at level
$n$ ($n \gg 1$) for fundamental $p-$branes compactified on manifold
with topology ${\mathbb T}^p\times {\mathbb R}^{d-p}$ has been
calculated within the semiclassical quantization scheme in
\cite{Bytsenko93,Bytsenko96}. In string theory, in the case of zero
modes, the embedding spacetime dependence can be eliminate
\cite{Rama}, the coefficient of logarithmic correction ${\mathcal
A}(p)$ becomes $-3/2$, which agrees with the results obtained in the
spin network formalism.

To summarize, our results can be used in the context of the brane
method to calculate  the ground state degeneracy of systems with
quantum numbers of certain BPS extreme black holes. For example, the
BPS black hole in toroidally compactified type II string theory. One
can construct a brane configuration such that the corresponding
supergravity solutions describe five-dimensional black holes. Black
holes in these theories can carry both an electric charge $Q_{F}$and
an axion charge $Q_{H}$. The brane picture gives the entropy in
terms of partition function $W(t)$ for a gas of $Q_{F}Q_{H}$ species
of massless quanta: $ W(t)=\prod_{{\bf n}\in {\mathbb Z}^{p}/\{{\bf
0}\}}\left[ 1-\exp \left( -t\omega _{{\bf n}}({\bf a},{\bf
g})\right)\right] ^{-\sigma} $, where $\sigma = ({\rm dim}\,X -
p-1)$,\, $t=y+2\pi ix$, $\Re\, t>0$. For unitary conformal theories
of fixed central charge $c$ the entropy becomes: $ {S}(n)={\rm log}
\chi (n)\simeq {S}_{0} + {\mathcal A}{\rm log} ({S}_{0}), $ where $
{S}_{0}=2\pi \sqrt{cn/6}\,, {\mathcal A} = - (c+3)/2. $ Following
\cite{Strominger1,Abdalla031}, we can put $c=3Q_{F}^{2}+6,\,\,
n=Q_{H}$ and get the growth of the elliptic genus (or the degeneracy
of BPS solitons) for $n=Q_{H}\gg 1$. However, this result is
incorrect when the black hole becomes massive enough and its
Schwarzschild radius exceed any microscopic scale such as the
compactification radii \cite{Maldacena,Halyo}. Such models, stemming
from string theory, would therefore be incompatible; in view of the
present result, this might be presented as a useful constraint for
the underlying microscopic field theory.

\section*{Acknowledgements}
The authors would like to thank the Conselho Nacional de
Desenvolvimento Cient\'{\i}fico e Tecnol\'ogico (CNPq, Brazil) and
Arauc\'aria foundation (Paran\'a-Brazil) for partial support.

\end{document}